\documentclass[a4paper,12pt]{article}

\usepackage[utf8]{inputenc}
\usepackage[T1]{fontenc}
\usepackage{amsmath, amssymb}
\usepackage{graphicx}
\usepackage{booktabs}
\usepackage{multirow}
\usepackage{geometry}
\usepackage{float}
\usepackage{hyperref}

\usepackage[backend=biber, style=numeric, maxbibnames=4, minbibnames=1]{biblatex}
\newcommand{\keywords}[1]{\noindent\textbf{Keywords:} #1\par}
\newcommand{\acknowledgments}{\section*{Acknowledgments}}
\newcommand{\conflictsofinterest}[1]{\section*{Conflicts of Interest} #1}
\newcommand{\funding}[1]{\section*{Funding} #1}
\newcommand{\institutionalreview}[1]{\section*{Institutional Review Board Statement} #1}
\newcommand{\informedconsent}[1]{\section*{Informed Consent Statement} #1}
\newcommand{\dataavailability}[1]{\section*{Data Availability Statement} #1}

\title{Methods for Centrality Determination Using Forward Detectors in the BM@N Experiment}

\author{
    Dim Idrisov$^{1,*}$,
    Fedor Guber$^{1,*}$,
    Nikolay Karpushkin$^{1}$,
    Peter Parfenov$^{1}$
}

\date{}

\begin{document}

\maketitle

\noindent
$^{1}$ Institute for Nuclear Research of the Russian Academy of Sciences, Prospekt 60-letiya Oktyabrya 7a, Moscow 117312, Russia\\
$^{*}$ Correspondence: idrisov.dim@mail.ru (D.I.); guber@inr.ru (F.G.)

\vspace{2ex}

\begin{abstract}
Collision centrality is a key parameter for studying nuclear matter properties, as it determines the initial interaction geometry and the size of the produced system. Accurate centrality determination is essential for comparing experimental data obtained from different experiments and for benchmarking against theoretical models. 
This work presents a modification of the approach for centrality determination using charged particle multiplicity based on Bayes' theorem. The proposed improvements enable an estimation of event registration efficiency as a function of the impact parameter. Furthermore, two approaches utilizing forward detectors are proposed: a two-dimensional method based on the combined analysis of track hit counts and spectator deposited energy in the Forward Hadron Calorimeter FHCal, and a method employing signals from the quartz hodoscope and FHCal. These methods were applied to data from the first physics run Xe+CsI of the BM@N experiment (Baryonic Matter at Nuclotron) with a xenon beam at the energy of 3.8 A GeV. A comparison of the developed methods with the classical Monte Carlo Glauber approach demonstrates agreement within 5\% across all considered methods, confirming their reliability and mutual consistency.
The use of forward detectors for centrality determination may serve as an independent tool for assessing the initial collision geometry and can reduce autocorrelation effects in studies of proton multiplicity fluctuations. The developed approaches can be employed for data processing in the BM@N experiment, as well as in other heavy-ion experiments at intermediate energies.
\end{abstract}


\section{Introduction}

The study of strongly interacting nuclear matter at high baryon densities is a central challenge in modern high-energy physics. Of particular interest is the region of the quantum chromodynamics (QCD) phase diagram characterized by a large baryon chemical potential, where phase transitions and critical phenomena are expected to occur. 

A crucial parameter necessary for the analysis and interpretation of data in heavy-ion collisions is collision centrality. Centrality characterizes the initial collision geometry and, consequently, important quantities such as the number of participants (\(N_{\text{part}}\)) and the number of binary nucleon-nucleon collisions (\(N_{\text{coll}}\)). Collective observables, strange particle yields, and fluctuations of conserved quantities directly depend on centrality. Centrality is directly related to the impact parameter \(b\), defined as the distance between the centers of the two colliding nuclei in the transverse plane. However, the impact parameter cannot be measured directly in an experiment; instead, the correlation between measured observables and the impact parameter is exploited \parencite{broniowski2002, alice2013}. Typically, charged-particle multiplicity in the central rapidity region serves as the observable of choice. An alternative approach involves the use of forward detectors that register spectator fragments, providing independent information on the collision geometry \parencite{kagamaster2021,idrisov_centrality_2024,segal_possibilities_2023}.

At low energies, the standard Glauber model faces several limitations: it does not account for Coulomb scattering on the target nucleus, baryon stopping effects, and associated energy losses in successive collisions \parencite{simak2025}, leading to violations of energy conservation. The low and discrete multiplicity of produced particles leads to significant statistical fluctuations, which complicates the division into narrow centrality classes. In the HADES experiment, the introduction of phenomenological parameterizations \parencite{hades_collaboration_centrality_2018} was required, likely indicating insufficient accuracy of the Kharzeev–Nardy approach in this energy region. Volume fluctuations are difficult to disentangle from dynamical ones, and for certain observables, such as net-proton multiplicity fluctuations, autocorrelation effects \parencite{luo_volume_2013,adamczewski-musch_proton-number_2020}, related to the use of the same multiplicity for the definition of centrality, play a significant role. The combination of these factors motivates the development of new, independent approaches for centrality estimation in experiment.

The energy range $\sqrt{s_\mathrm{NN}} = 2.4 - 11$~GeV represents one of the most promising domains in the field of heavy-ion collisions. In this regime, the existence of the critical point and the onset of a first-order phase transition are anticipated. Currently, several experiments are actively exploring this region, including NA61/SHINE at the SPS, STAR at RHIC, and BM@N (Baryonic Matter at Nuclotron) at NICA~\parencite{afanasiev2024}, as well as upcoming facilities such as CBM at SIS100, CEE at HIRFL-CSR, and the MPD (Multi-Purpose Detector) at NICA. In this work, we focus on the BM@N experiment to test the centrality determination methods developed in this study. The primary objective of BM@N is to investigate the properties of dense baryonic matter formed in nucleus-nucleus collisions at beam energies ranging from 1 to 4.5A~GeV~\parencite{senger2022, BMAN2023}. In this energy region, baryon densities of up to 3–4 times normal nuclear density are achieved, enabling detailed studies of the nuclear matter equation of state, mechanisms of strange particle production, and collective effects~\parencite{afanasiev2024}.

This work presents the results of developing and validating centrality determination methods in the BM@N experiment. The classical Monte Carlo Glauber method \parencite{miller2007} and the one-dimensional direct reconstruction method (\(\Gamma\)-fit) based on Bayes' theorem \parencite{das2018, rogly2018, indra2020}, both utilizing charged particle multiplicity, are considered, alongside a newly developed two-dimensional method that combines track hit counts with spectator energy measured by the Forward Hadron Calorimeter FHCal. A novel approach for centrality determination using signals from the quartz hodoscope and FHCal is proposed. These methods were applied to data from the first physics run with a xenon beam Xe+CsI at an energy of 3.8 A GeV.

\section{BM@N Detectors}

\begin{figure}[H]
\centering
\includegraphics[width=0.75\textwidth]{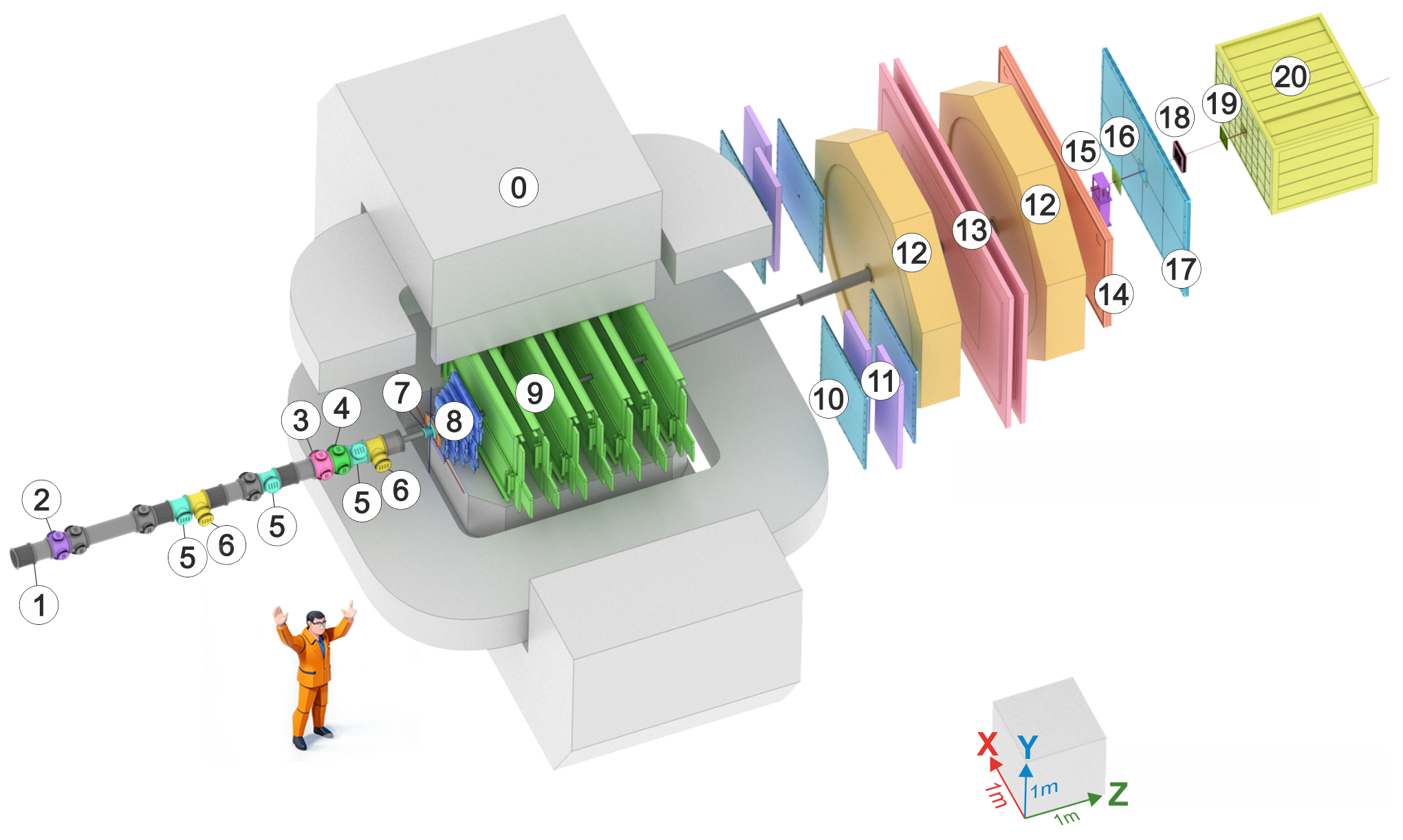}
\caption{Schematic view of the BM@N setup in the 2023 Xe run. Main components: 0) SP-41 analyzing magnet, 1) vacuum beam pipe, 2-4) beam counters, 5) Silicon Beam Tracker (SiBT), 6) Silicon beam profilometers, 7) Barrel Detector and Target station, 8) Forward Silicon Detector, 9) Gaseous Electron Multiplier detectors, 10) Small cathode strip chambers, 11) time-of-flight system TOF400, 12) drift chambers, 13) time-of-flight system TOF700, 14) Scintillation Wall, 15) Fragment Detector, 16) Small GEM detector, 17) Large cathode strip chamber, 18) gas ionization chamber as beam profilometer, 19) Forward Quartz Hodoscope, 20) Forward Hadron Calorimeter. \parencite{afanasiev2024}.}
\label{fig:setup}
\end{figure}

The BM@N experimental setup is a large-acceptance magnetic spectrometer operating with a fixed target at the extracted beams of the NICA complex \parencite{afanasiev2024}. The BM@N detector configuration comprises a set of beam and trigger detectors, coordinate and time-of-flight detectors for measuring the momenta of produced particles and their identification, as well as a forward hadron calorimeter and a quartz hodoscope (Fig. \ref{fig:setup}) \parencite{afanasiev2024}.

The BM@N trigger system is based on fast scintillation counters and generates the signal to initiate event readout. It primarily consists of beam counters BC1 and BC2, which detect the passage of the ion, and a veto counter VC with a 25 mm aperture, which rejects background events occurring outside the target. The Barrel Detector BD, surrounding the target, measures charged particle multiplicity and is used to select central collisions (CCT trigger). The Fragment Detector FD, located downstream of the magnet, measures the ionization losses of projectile fragments, allowing in-target interaction events to be distinguished from beam passage (MBT trigger). The combination of these signals ensures efficient selection of events with varying centrality.

The BM@N spectrometer tracking system includes silicon and gaseous detectors positioned within the aperture of the analyzing magnet SP-41. The Forward Silicon Detector FSD, installed directly downstream of the target, consists of four planes of double-sided silicon strip sensors and provides precise coordinate measurements of secondary particles near the interaction point. The Gas Electron Multiplier system, comprising seven coordinate planes located above and below the beam pipe, allows reconstruction of particle trajectories in the magnetic field with high spatial resolution. The combined use of FSD and GEM detector data enables momentum reconstruction of secondary charged particles over a wide pseudorapidity range. The number of hits recorded in the tracking system is used as one of the observables in the two-dimensional method.

The Forward Hadron Calorimeter FHCal is installed at the downstream end of the spectrometer to measure spectator fragment energy and determine collision centrality. The calorimeter consists of 54 sandwich-type modules (lead/scintillator) with longitudinal segmentation in a compensated configuration, achieving a linear response over a wide energy range. A square aperture in the central part of the calorimeter allows the passage of non-interacting beam ions.

The Forward Quartz Hodoscope FQH is used to register the charge of projectile nucleus fragments traversing the FHCal aperture. The detector consists of 16 quartz strips. Combined analysis of data from the FHCal and FQH enables the reconstruction of the impact parameter across the entire centrality range.

\section{Centrality Estimation in Nucleus-Nucleus Collisions from Produced Particle Multiplicity}

This section describes the methods for centrality determination using produced particle multiplicity in nucleus-nucleus collisions as a more conventional methods. The first approach is based an implementation of the Monte Carlo version of the Glauber model described in \parencite{loizides2015,demanov_methods_2025,parfenov2021}.

The input parameter in the Glauber model is the nucleon density \(\rho(r)\) inside the nucleus, which can be described by a Fermi distribution:
\begin{equation}
\label{eq:McGlauberFermi}
\rho(r) = \rho_0 \frac{1}{1+\exp{\frac{r-R}{a}}},
\end{equation}
where \(R\) is the nuclear radius (\(R_{Xe} = 5.42 \pm 0.05\) fm for the \(^{124}\text{Xe}\) nucleus), the constant \(\rho_0\) corresponds to the density at the nuclear center, and the parameter \(a = 0.54 \pm 0.01\) fm characterizes the surface thickness.

A nucleus-nucleus collision is treated as a sequence of independent binary nucleon-nucleon collisions, where nucleons travel along straight-line trajectories. The inelastic nucleon-nucleon cross-section \(\sigma_{\text{NN}}^{\text{inel}}\) is assumed to depend only on the collision energy. Two nucleons from different nuclei interact if the relative transverse distance \(d\) between their centers satisfies \(d < \sqrt{\sigma_{\text{NN}}^{\text{inel}}/\pi}\). In this work, a value of \(\sigma_{\text{NN}}^{\text{inel}} = 29.4\) mb, corresponding to an energy of 3.8 A GeV, was used \parencite{ParticleDataGroup:2024cfk}.

The outputs of the MC Glauber model include the geometric properties of the simulated collisions: impact parameter \(b\), number of binary collisions (\(N_{\text{coll}}\)), and number of participant nucleons (\(N_{\text{part}}\)). The centrality determination procedure consists of fitting the function \(N_{ch}^{\text{fit}}(f, \mu, k)\) from the MC Glauber model to the experimentally measured charged particle multiplicity \(N_{ch}\):
\begin{equation}
N_{ch}^{\text{fit}}(f, \mu, k) = N_a(f) \times P_{\mu,k},\quad N_a(f) = fN_{\text{part}} + (1-f)N_{\text{coll}},
\end{equation}
where \(P_{\mu,k}\) is the negative binomial distribution (NBD) with mean \(\mu\) and parameter \(k\). The quantity \(N_{a}(f)\) represents the number of independent sources (ancestors), and the parameter \(f\) characterizes the fraction of hard processes. The optimal set of parameters \(f\), \(\mu\), and \(k\) is found by minimizing \(\chi^2\):
\begin{equation}
\label{eq:McGlauberChi2}
\chi^2 = \sum \limits_{i=n_{\text{low}}}^{n_{\text{high}}} \frac{\left( F_{\text{fit}}^i - F_{\text{data}}^i \right)^2}{\left( \Delta F_{\text{fit}}^i \right)^2 + \left( \Delta F_{\text{data}}^i \right)^2},
\end{equation}
where \(F_{\text{fit}}^i\) and \(F_{\text{data}}^i\) are the values of the fitting function and the data histogram in specified intervals \(i\), \(\Delta F_{\text{fit}}^i\) and \(\Delta F_{\text{data}}^i\) are the corresponding uncertainties, and \(n_{\text{low}}\) and \(n_{\text{high}}\) are the lower and upper fit boundaries, respectively.

As energy decreases, 
the Coulomb interactions become non-negligible \parencite{mehndiratta2017} and energy conservation constraints in sequential nucleon interactions \parencite{simak2025} into the Glauber model grows. Despite these considerations, the Glauber model has been used without modifications down to 3 GeV \parencite{abdallah_measurements_2022}. However, at lower energies of 1.23A GeV, a new phenomenological parameterization \parencite{hades_collaboration_centrality_2018} was required to describe charged particle multiplicity, possibly indicating a change in collision dynamics in this energy regime.

\begin{figure}[H]
\centering
\includegraphics[width=0.6\textwidth]{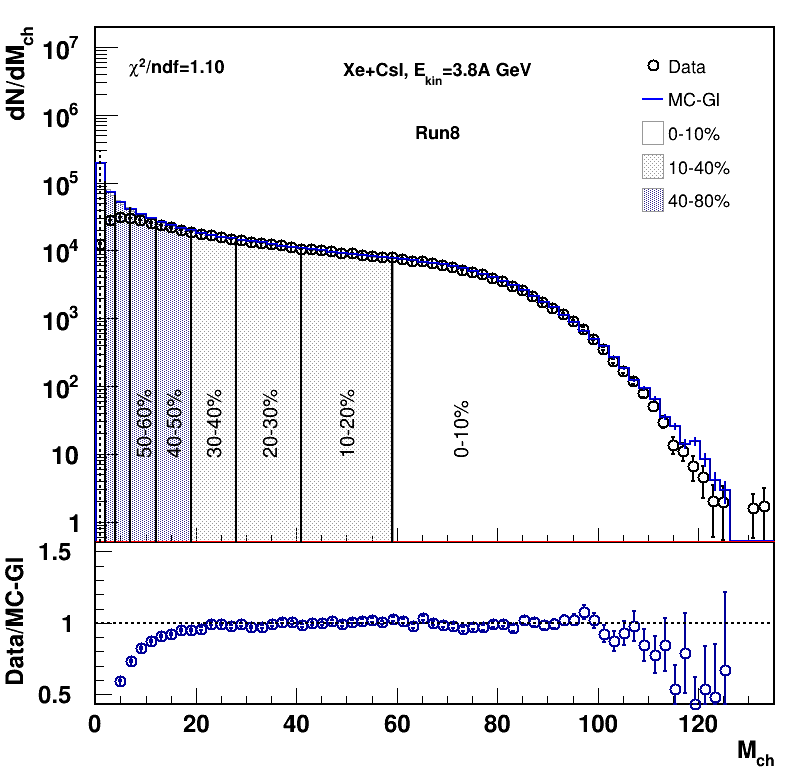}
\caption{Results of fitting the charged particle multiplicity using the Monte Carlo Glauber method. Blue lines represent the fit result, open symbols show the experimental multiplicity distribution.}
\label{fig:MCGlauber}
\end{figure}

Figure \ref{fig:MCGlauber} shows the results of fitting the charged particle multiplicity for the first BM@N physics run for the Xe+CsI reaction at 3.8 A GeV using the Monte Carlo Glauber method. The bottom panel illustrate the ratio of experimental data to the fit. To suppress contributions from secondary pions and spectators to the charged particle multiplicity, only tracks satisfying the following conditions were selected: |DCA|<1 cm, 0.05 < $Pt$<2 GeV/c, 0.7<$\eta$<2.7. The results indicate that the Glauber model reproduces the charged particle multiplicity rather well. The discrepancy between the fit and data in the peripheral region is attributed to the event registration efficiency, which is determined by the trigger system and event selection criteria. The onset of this deviation is conventionally defined as the ''anchor point''. Although a detailed characterization of the anchor point is beyond the scope of the present study, it remains a critical parameter for subsequent experimental analyzes. This divergence indicates that the trigger system introduces a potential bias for observables measured in the events beyond the anchor point.

The second approach used for centrality determination is the direct reconstruction method based on Bayes' theorem \parencite{das2018, rogly2018}. It allows extracting information about the impact parameter \(b\) using only the measured distribution of an observable \(X\), without simulating the entire collision process. Any additive quantity correlated with \(b\) can serve as the observable \(X\). This method has been tested on data from the ALICE and STAR experiments \parencite{das2018, rogly2018} and has also demonstrated applicability at low energies in the INDRA experiment \parencite{indra2020}.

The charged particle multiplicity fluctuations can be described by a gamma distribution \parencite{rogly2018}. For the charged particle multiplicity \(N_{ch}\), this distribution has the form:
\begin{equation}
\label{eq:GammaP}
P(N_{ch}|b) = \frac{1}{\Gamma(k)\theta^k} N_{ch}^{k-1} e^{-N_{ch}/\theta},
\end{equation}
where \(\Gamma(k)\) is the gamma function, and \(k\) and \(\theta\) are distribution parameters corresponding to the mean \(\langle N_{ch} \rangle\) and the standard deviation \(\sigma_{N_{ch}}\):
\begin{equation}
k=\frac{\langle N_{ch} \rangle^{2}}{\sigma_{N_{ch}}^{2}}, \quad \theta=\frac{\sigma_{N_{ch}}^{2}}{\langle N_{ch}\rangle}.
\label{eq:GammaPar}
\end{equation}
Thus, the gamma distribution parameters \(k\) and \(\theta\) are smooth continuous functions depending on the impact parameter \(b\) or centrality \(c_{b}\). Like the charged particle multiplicity \(N_{ch}\), the gamma distribution is defined only for \(N_{ch} \geq 0\). It can be regarded as a continuous counterpart of the negative binomial distribution (NBD).

The normalized measured multiplicity distribution \(P(N_{ch})\) can be obtained by summing contributions over all impact parameters:
\begin{equation}
\label{convolution}
P(N_{ch})=\int_0^{\infty} P(N_{ch}|b)P(b)db=\int_0^1 P(N_{ch}|c_b)dc_b,\quad P(b) = \frac{2\pi b}{\sigma_{\text{inel}}} P_{\text{inel}}(b),
\end{equation}
where \(P(b)\) is the probability density distribution of the impact parameter, depending on the inelastic collision probability \(P_{\text{inel}}(b)\) for a given \(b\) and the inelastic nucleus-nucleus interaction cross-section \(\sigma_{\text{inel}}\). The variable \(c_{b}\) denotes centrality defined via the impact parameter: \(c_b =\int_0^b P(b')db'\).

Thus, in proposed modification of the direct reconstruction method, the main distinction is to the use of distributions of the observable mean and its variance derived from the model. The advantage of this approach is a more realistic approximation of \(P(b)\) from the model and the ability to incorporate calculated efficiency, allowing more precise centrality determination in experimental data. Here, ''calculated efficiency'' refers to the combined efficiency resulting from the trigger system efficiency and the event selection conditions.

To account for the experimental effects on the charged particle multiplicity it is parametrized as a function of two random variables: the number of particles per source \(n_i\) and the number of sources \(N_a\): \(N_{ch} =\sum_{i=1}^{N_{a}} n_{i}\). Then, according to the law of total variance, the variance of the charged particle multiplicity is:
\begin{equation}
\sigma_{N_{ch}}^{2} = \langle N_{\text{a}} \rangle \sigma_n^2 + \langle n \rangle^2 \sigma_{N_{\text{a}}}^{2},
\end{equation}
where \(\sigma_n^2\) is the variance of the multiplicity per source. Assume that in the model, the number of sources \(N_{\text{a}}\approx N_{\text{a}}^{\text{MC}}\) is reproduced sufficiently well, but the distribution \(n^{\text{MC}}\) differs from the experimental one due to detector efficiency, cross-section differences, etc. Let \(n = \alpha_{N} n^{\text{MC}}\), then \(\langle N_{ch} \rangle = \alpha_{N} \langle N_{ch}^{\text{MC}} \rangle\). The relationship between the experimental and model variances is:
\begin{equation}
\sigma_{N_{ch}}^{2} = \alpha_N \beta_N \langle N_{ch}^{\text{MC}} \rangle + \alpha_N^{2} \sigma_{N_{ch}^{\text{MC}}}^{2},
\end{equation}
where \(\beta_N = (\sigma_{n}^{2} - \alpha_N^{2} \sigma_{n^{\text{MC}}}^{2}) / \langle n^{\text{MC}} \rangle\). Obtaining the distributions of the mean \(\langle N_{ch}^{\text{MC}}(c_{b}) \rangle\) and variance \(\sigma_{N_{ch}^{\text{MC}}}^{2}(c_{b})\) from the model and introducing two constants \(\alpha_N, \beta_N\), one can approximate the experimental multiplicity means and variances. Using this approximation, \(k(c_b)\) and \(\theta(c_b)\) are calculated according to \eqref{eq:GammaPar}.

\begin{figure}[H]
\centering
\includegraphics[width=0.6\linewidth]{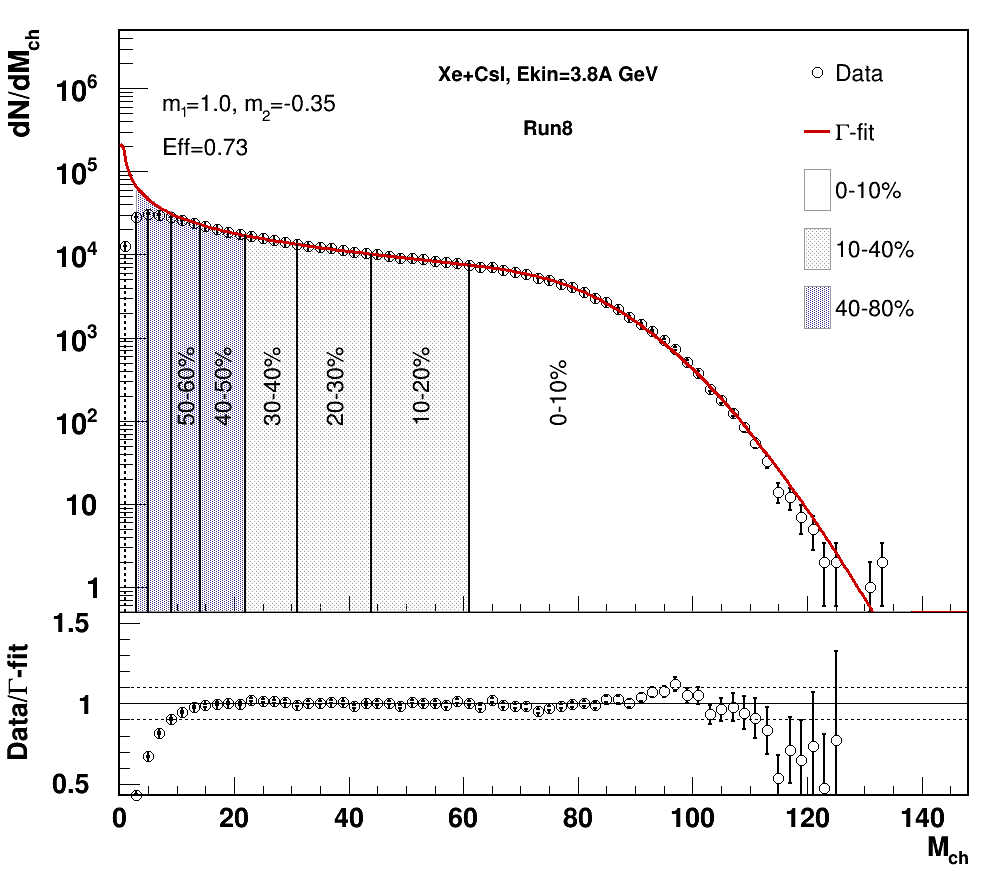}
\caption{Results of fitting the charged particle multiplicity using the direct reconstruction method. The red line shows the fit function, open symbols represent the experimental multiplicity distribution.}
\label{fig:FitGamma}
\end{figure}

Another important factor to consider is the calculated efficiency. Due to its finite efficiency, a fraction of events is lost, affecting the distribution normalization. To account for this effect, a parameter \(\epsilon = N_{\text{Raw}}^{Ev}/N_{\text{Total}}^{Ev}\) is introduced, where \(N_{\text{Raw}}^{Ev}\) is the number of events registered in the experiment and \(N_{\text{Total}}^{Ev}\) is the total number of inelastic events, corresponding to events that would be registered by an ideal trigger. The fit function for the normalized experimental multiplicity distribution is:
\begin{equation}
F(N_{ch}) = \frac{1}{\epsilon} \int_0^1 P(N_{ch}|c_{b}) dc_{b}.
\label{eq:1DFitFunc}
\end{equation}
Figure \ref{fig:FitGamma} presents the fit results for the multiplicity distribution for Xe+CsI collisions at 3.8 A GeV in the BM@N experiment. 

\begin{figure}[H]
\centering
\includegraphics[width=0.6\linewidth]{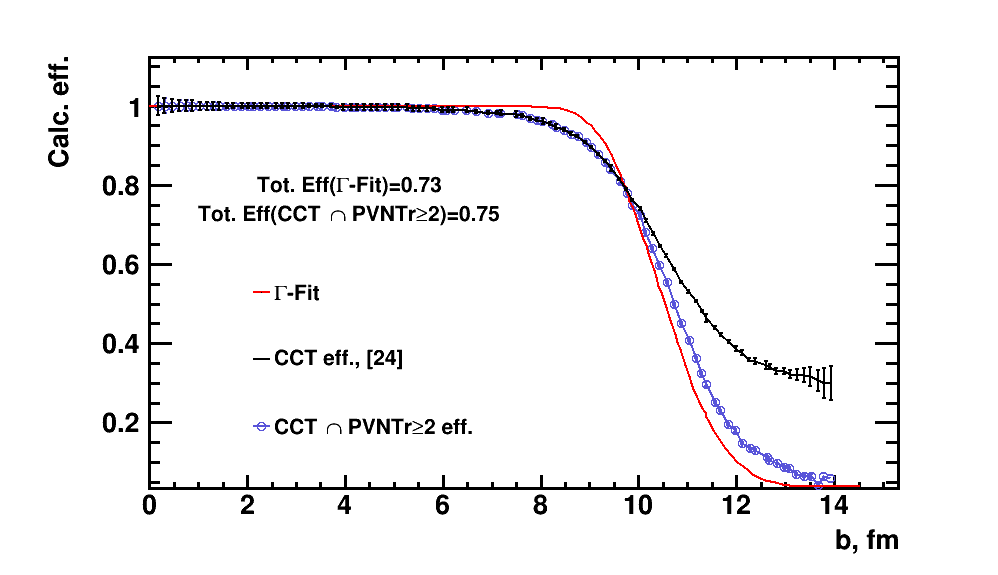}
\caption{Calculated efficiency as a function of impact parameter. }
\label{fig:eff}
\end{figure}

The red line represents function \eqref{eq:1DFitFunc}, and the bottom panel shows the ratio of experimental data to the fit function. The "anchor point" value is similar to the one obtained from Monte-Carlo Glauber results. This distribution can be interpreted as the calculated efficiency as a function of multiplicity \(F_{\text{eff}}(N_{ch})\). Using this distribution and the relationship between multiplicity and impact parameter the calculated efficiency as a function of impact parameter can be obtained:
\begin{equation}
F_{\text{eff}}(b) = \int_0^{\infty} F_{\text{eff}}(N_{ch}) P(N_{ch}|b) dN_{ch}.
\label{eq:1DFitEff}
\end{equation}

The efficiency calculated according to \eqref{eq:1DFitEff} for the experimental data is shown in Figure \ref{fig:eff} (red line). The black line shows the efficiency from Ref. \parencite{plotnikov2025} for the CCT trigger calculated based on experimental data. 

The dependence of the efficiency on the impact parameter was obtained by mapping the efficiency dependence on primary vertex multiplicity onto the distributions from the DCM-SMM model. To correct comparison we accounted for the vertex reconstruction efficiency (\(PVNTr \ge 2\)) for the results from Ref. \parencite{plotnikov2025}. The obtained results are shown in Figure \ref{fig:eff} as open blue symbols. The recalculated CCT efficiency, taking into account the vertex registration efficiency, is in good agreement with the results of the direct reconstruction method; the total efficiency agrees within 2\%. The obtained dependence allows correcting model distributions when comparing with experimental data and accounts for event losses in the peripheral region.

\section{Two-Dimensional Direct Reconstruction Method}

To address the issues with the multiplicity-based approach mentioned above, we introduce the two-dimensional method for centrality determination. A central idea of the proposed new method \parencite{idrisov2025twodimensionalbayesianapproachcentrality} is that the function \(P(E_{\text{spec}}, N_{\text{hits}}|b)\), characterizing the probability density distribution of spectator energy \(E_{\text{spec}}\) deposited in the forward hadron calorimeter and track hit count \(N_{\text{hits}}\) at fixed \(b\), can be approximated as:

\begin{equation}
\begin{gathered} 
P(E_{\text{spec}}, N_{\text{hits}}|b)=P(X_1(E_{\text{spec}}, N_{\text{hits}}), X_2(E_{\text{spec}}, N_{\text{hits}})|b)\\ = \prod_{i=1}^2 \frac{X_i^{k_i(b)-1} e^{-X_i/\theta_i(b)}}{\Gamma(k_i(b)) \theta_i(b)^{k_i(b)}},
\end{gathered} 
\end{equation}
where \(k_i(b)\) and \(\theta_i(b)\) are positive parameters depending on \(b\) and defined as:
\begin{equation}
k_i(b) = \frac{\langle X_i(b) \rangle^2}{\sigma_{X_i}^2(b)}, \quad \theta_i(b) = \frac{\sigma_{X_i}^2(b)}{\langle X_i(b) \rangle}.
\end{equation}
The new variables \(X_1\) and \(X_2\) are obtained by suppressing the linear component of the correlation between \(E_{\text{spec}}\) and \(N_{\text{hits}}\). The simplest way to suppress such correlation is rotation by an angle:
\begin{equation}
\phi(b)=\frac{1}{2}\arctan\left( \frac{2\,\text{Cov}(E_{\text{spec}}, N_{\text{hits}})}{\sigma_{E_{\text{spec}}}^2-\sigma_{N_{\text{hits}}}^2} \right).
\end{equation}
The rotation angle is calculated from the condition that the covariance of the new variables vanishes \(\text{Cov}(X_1, X_2) = 0\). The relationship between \(X_1\), \(X_2\) and \(N_{\text{hits}}\), \(E_{\text{spec}}\) is defined via a rotation matrix, and the means and variances are:
\begin{equation}
\begin{pmatrix}
\langle X_1 \rangle\\
\langle X_2 \rangle
\end{pmatrix}
=
\begin{pmatrix}
\cos\phi & -\sin\phi \\
\sin\phi & \cos\phi
\end{pmatrix}
\begin{pmatrix}
\langle N_{\text{hits}} \rangle\\
\langle E_{\text{spec}} \rangle
\end{pmatrix},
\end{equation}
\begin{equation}
\sigma_{X_1}^2 = \cos^2\phi \sigma_{N_{\text{hits}}}^2 + \sin^2\phi \sigma_{E_{\text{spec}}}^2 - \sin(2\phi) \text{Cov}(N_{\text{hits}}, E_{\text{spec}}),
\end{equation}
\begin{equation}
\sigma_{X_2}^2 = \sin^2\phi \sigma_{N_{\text{hits}}}^2 + \cos^2\phi \sigma_{E_{\text{spec}}}^2 + \sin(2\phi) \text{Cov}(N_{\text{hits}}, E_{\text{spec}}).
\end{equation}
The distributions of the mean, variance, and covariance of the observables from the model are approximated by polynomials \(f(c_b)\). Thus, the fit function for the spectator energy and track hit count distribution \(F(E_{\text{spec}}, N_{\text{hits}})\) is related to the probability density at fixed impact parameter \(P(E_{\text{spec}}, N_{\text{hits}}|b)\) as:
\begin{equation}
F(E_{\text{spec}}, N_{\text{hits}}) = \frac{1}{\epsilon}\int_0^1 P(E_{\text{spec}}, N_{\text{hits}}|c_{b}) dc_{b}.
\label{eq:2DFit}
\end{equation}
Analogous to the one-dimensional approach, it is assumed that the spectator energy and track hit count in the experiment are proportional to the values obtained from fully reconstructed model data:
\begin{equation}
\langle E_{\text{spec}}(b) \rangle = \alpha_E \langle E_{\text{spec}}^{\text{MC}}(b)\rangle,\quad
\langle N_{\text{hits}}(b) \rangle = \alpha_N \langle N_{\text{hits}}^{\text{MC}}(b)\rangle,
\end{equation}
\begin{equation}
\sigma_{E_{\text{spec}}}^2(b) = \alpha_E^{2}\sigma_{E_{\text{spec}}^{\text{MC}}}^2(b)+\alpha_E\beta_E\langle E_{\text{spec}}^{\text{MC}}(b) \rangle,
\end{equation}
\begin{equation}
\sigma_{N_{\text{hits}}}^2(b) = \alpha_N^{2}\sigma_{N_{\text{hits}}^{\text{MC}}}^2(b)+\alpha_N\beta_N\langle N_{\text{hits}}^{\text{MC}}(b) \rangle.
\end{equation}
Thus, the method under consideration includes the following fit parameters: \(\alpha_E, \alpha_N, \beta_E, \beta_N, \epsilon\), which can be determined by fitting the two-dimensional distribution. The coefficients introduced in this way allow accounting for differences between experimental and model data in describing the means and variances of the observables. Such differences are due to variations in the gain coefficients of individual calorimeter readout channels and calibration uncertainties. Once the fit parameters are determined, the probability distribution of the impact parameter \(P(b|E_{\text{spec}}, N_{\text{hits}})\) for a fixed range of multiplicity (\(N_{1},\ N_{2}\)) and energy (\(E_{1},\ E_{2}\)) can be obtained from Bayes' theorem:
\begin{equation}
\begin{gathered} 
P(b|E_{1}<E_{\text{spec}}<E_{2},N_{1}<N_{\text{hits}}<N_{2})=\\
P_{\text{inel}}(b)\frac{\displaystyle\int_{E_{1}}^{E_{2}} \int_{N_{1}}^{N_{2}} P(E,N|c_{b})\,dE\,dN}{\displaystyle\int_{E_{1}}^{E_{2}} \int_{N_{1}}^{N_{2}} \int_{0}^{1} P(E,N|c_{b})\,dc_{b}\,dE\,dN}.
\end{gathered} 
\end{equation}

\section{Comparison of Centrality Determination Methods in the BM@N Experiment}

\begin{figure}[H]
\centering
\includegraphics[width=0.6\textwidth]{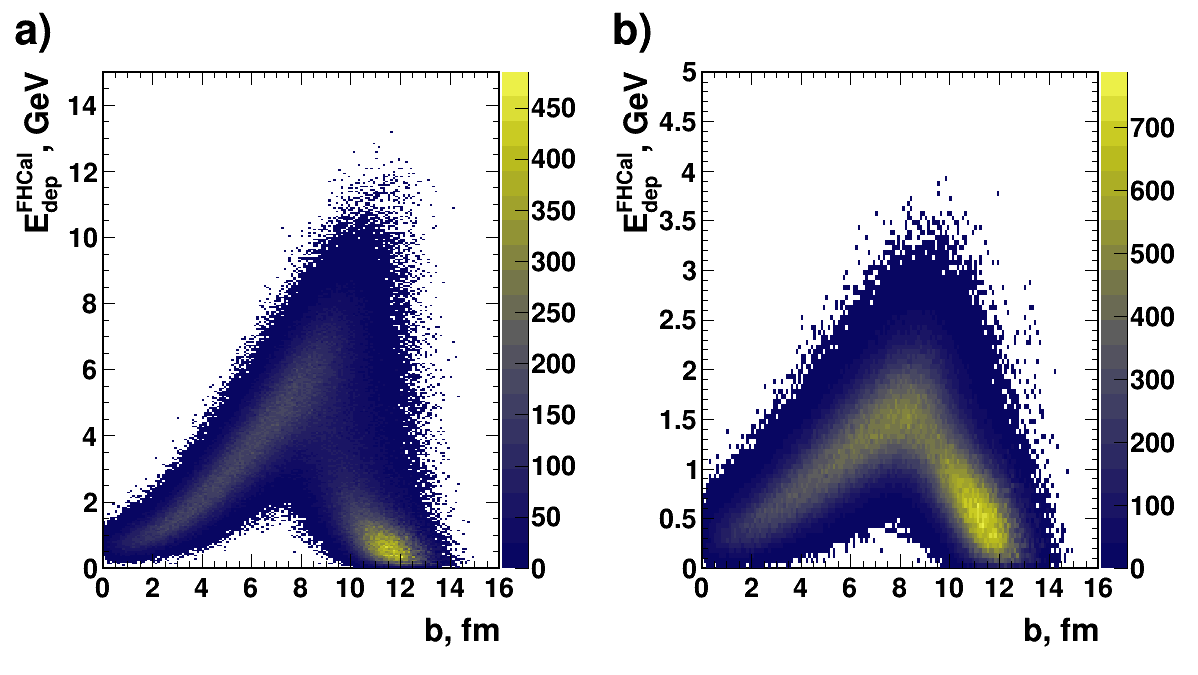}
\caption{Correlation between spectator energy deposited in FHCal and impact parameter for Xe+CsI collisions at beam energy 3.8 A GeV in the DCM-QGSM-SMM model: a) for all modules, b) for outer modules \parencite{afanasiev2024}.}
\label{fig:EnvsImp}
\end{figure}

To implement the two-dimensional method, the response characteristics of the FHCal calorimeter must be correctly accounted for. As shown in Figure \ref{fig:EnvsImp}, the correlation in simulated data between deposited energy and impact parameter differs for different groups of modules. 

When using the signal from all modules (Fig. \ref{fig:EnvsImp} left), the correlation with impact parameter becomes less monotonic. This effect is related to a specific feature of this run, as the beam direction was not coaxial with the detector axis. Furthermore, the energy distribution at a fixed impact parameter can no longer be described by a gamma distribution. Therefore, for the two-dimensional approach involving track hit counts, the energy deposited in the outer FHCal modules (Fig. \ref{fig:EnvsImp} right) was used, as it is sufficiently well described by a gamma distribution.

\begin{figure}[H]
\centering
\includegraphics[width=0.85\textwidth]{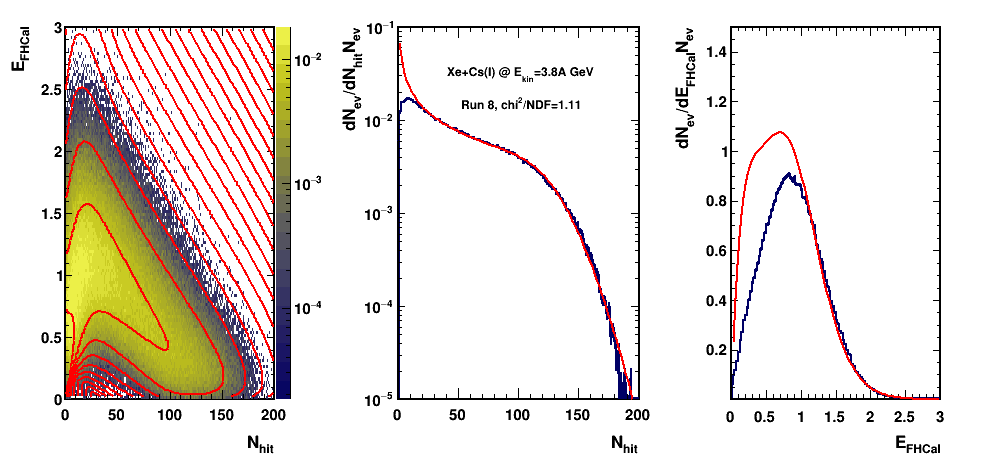}
\caption{(Left) Fit results for the two-dimensional distribution of hit counts and energy from the FHCal calorimeter. Red lines indicate contour lines of the fit function (\ref{eq:2DFit}). The central panel shows the projection onto the hit count axis, the right panel shows the projection onto the FHCal energy axis.}
\label{fig:2d_fit}
\end{figure}

Figure \ref{fig:2d_fit} shows the fit results for the two-dimensional distribution of spectator deposited energy and track hit count from experimental data for Xe+CsI collisions at 3.8 A GeV in the BM@N experiment. Function \eqref{eq:2DFit} including efficiency \(\epsilon\) describes the experimental data well across the entire observable range. Discrepancies between the fit and experimental data in the peripheral region for track hit counts are due to calculated efficiency (Fig. \ref{fig:2d_fit}). A similar situation is observed for spectator energy deposited in FHCal, but over a wider range of the distribution, since due to the absence of a central module, peripheral events correspond to a broad energy range.

\begin{figure}[H]
\centering
\includegraphics[width=0.5\textwidth]{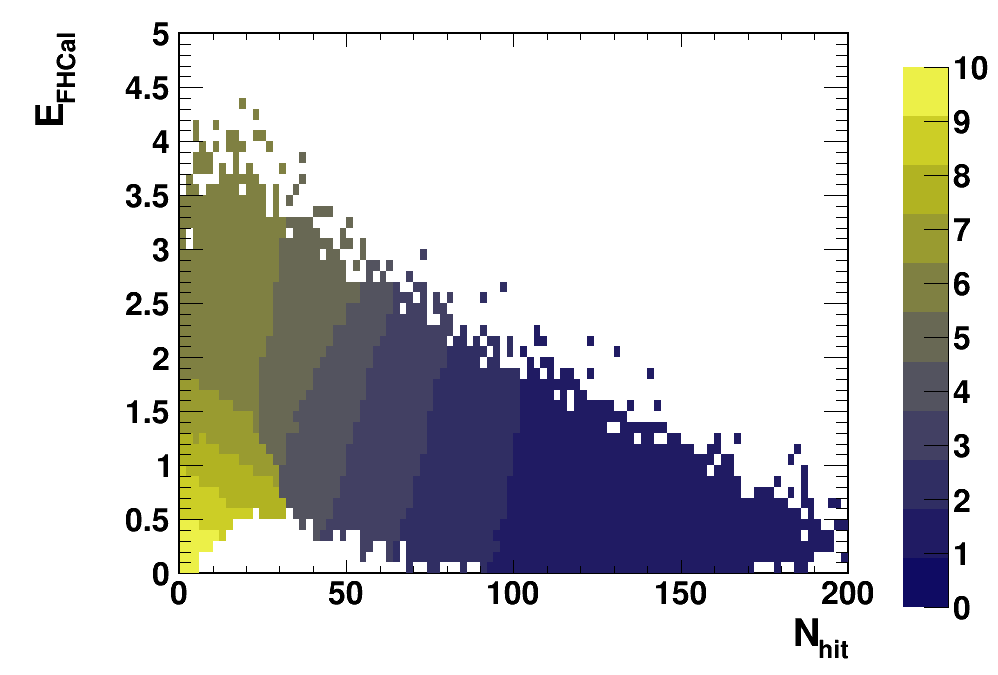}
\caption{Centrality classes for the two-dimensional distribution of track hit counts and FHCal energy.}
\label{fig:centrality_classes2D}
\end{figure}

After fitting, the obtained two-dimensional distribution was divided into 10 centrality classes using the k-means constrained method \parencite{ConstrainedKMeans} (Fig. \ref{fig:centrality_classes2D}). The mean observable values for the corresponding classes, obtained from previously determined centrality dependencies of the means, were used as cluster centers.

\begin{figure}[H]
\centering
\includegraphics[width=0.85\textwidth]{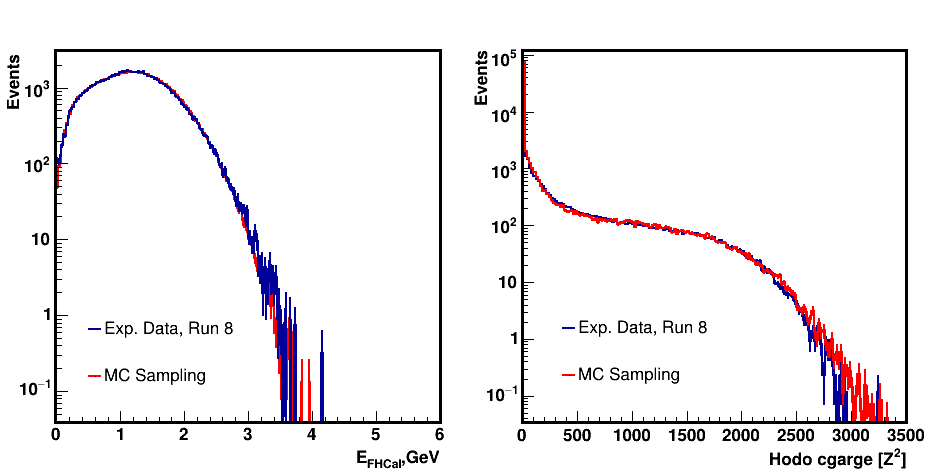}
\caption{Comparison of total energy in the calorimeter from model and experiment after scaling.}
\label{fig:FitHodo}
\end{figure}

After determining the coefficients \(\alpha_E,\beta_E\) from the two-dimensional fit, the total energy from the outer and central modules in the calorimeter can be compared between the DCM-QGSM-SMM model and the experiment. The results of this comparison, accounting for calculated efficiency obtained from the one-dimensional fit, are shown in Figure \ref{fig:FitHodo} (left). The obtained results demonstrate good agreement between the DCM-QGSM-SMM model data and the experiment.

\begin{figure}[H]
\centering
\includegraphics[width=0.5\textwidth]{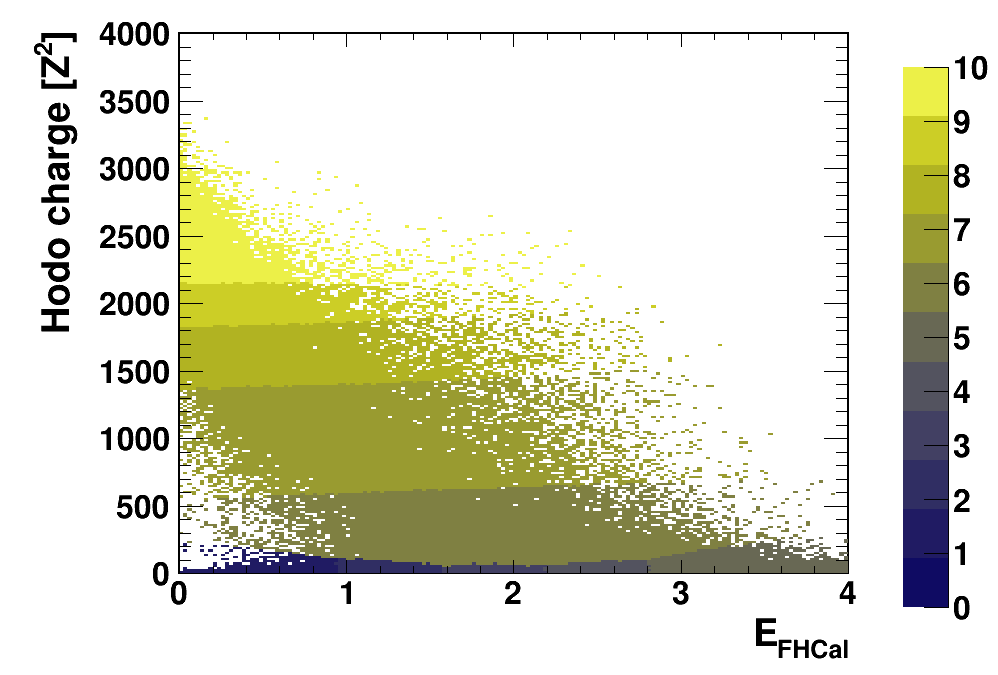}
\caption{Centrality classes for the two-dimensional distribution of signals from the quartz hodoscope and forward hadron calorimeter.}
\label{fig:centrality_classesHodo}
\end{figure}

Using the impact parameter dependence of efficiency \eqref{eq:1DFitEff} and the fit parameters \(\alpha_E,\beta_E\), another approach for centrality determination was developed. This approach uses signals exclusively from the forward detectors FHCal and FQH. Its advantage is low sensitivity to pile-up events due to a narrower time window compared to the tracking system, as well as the ability to select centrality classes for observables sensitive to autocorrelation effects. This also required fitting the charge-squared distribution in the quartz hodoscope. The fitting method is similar to that used in the Monte Carlo Glauber method, but instead of sampling the number of sources and impact parameter, a full model with fragmentation DCM-QGSM-SMM \parencite{baznat2020} and detector response simulation in GEANT4 \parencite{Geant4} is employed. The model signal is smeared with a Gaussian kernel to simulate electronic response; the smearing coefficients are determined by minimizing \(\chi^2/\text{NDF}\). The fit results for the hodoscope signal are shown in Figure \ref{fig:FitHodo} (right).

After matching the signals from the forward detectors in the model and experiment, the scaled minimum bias model data are divided into 10 centrality classes using the k-means constrained method. As in the previous approach, the number of events in each class is equal, and the cluster centers are chosen based on the mean observable values. Figure \ref{fig:centrality_classesHodo} shows the obtained centrality classes for the two-dimensional distribution of signals from the quartz hodoscope and FHCal.

\begin{figure}[H]
\centering
\includegraphics[width=0.5\textwidth]{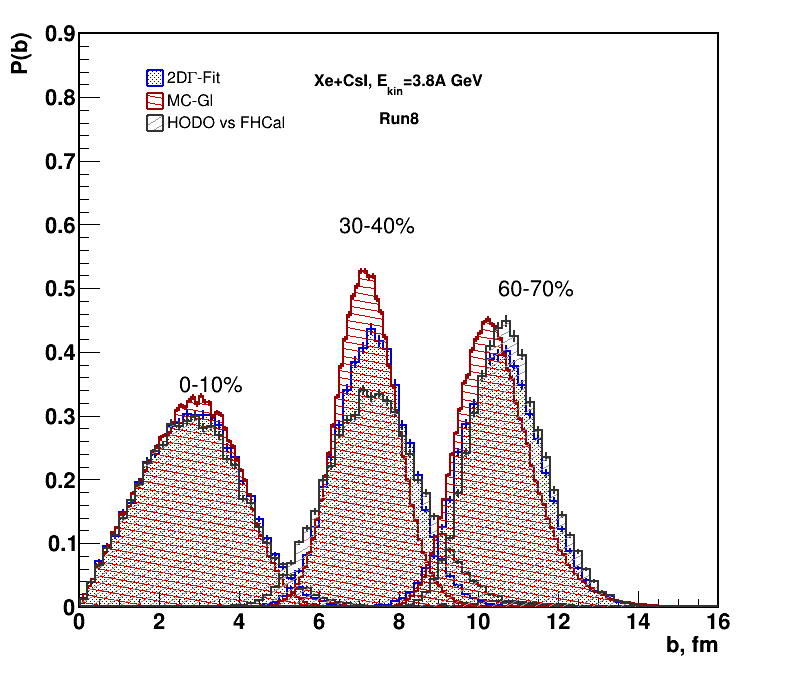}
\caption{Impact parameter distributions for three centrality classes(0-10\%, 30-40\%, and 60-70\%) obtained by MC-Glauber, two-dimensional methods with (\(E_{\text{spec}}\), \(N_{\text{hits}}\)) and only signals from the forward detectors.}
\label{fig:impDist}
\end{figure}

\begin{figure}[H]
\centering
\includegraphics[width=0.5\textwidth]{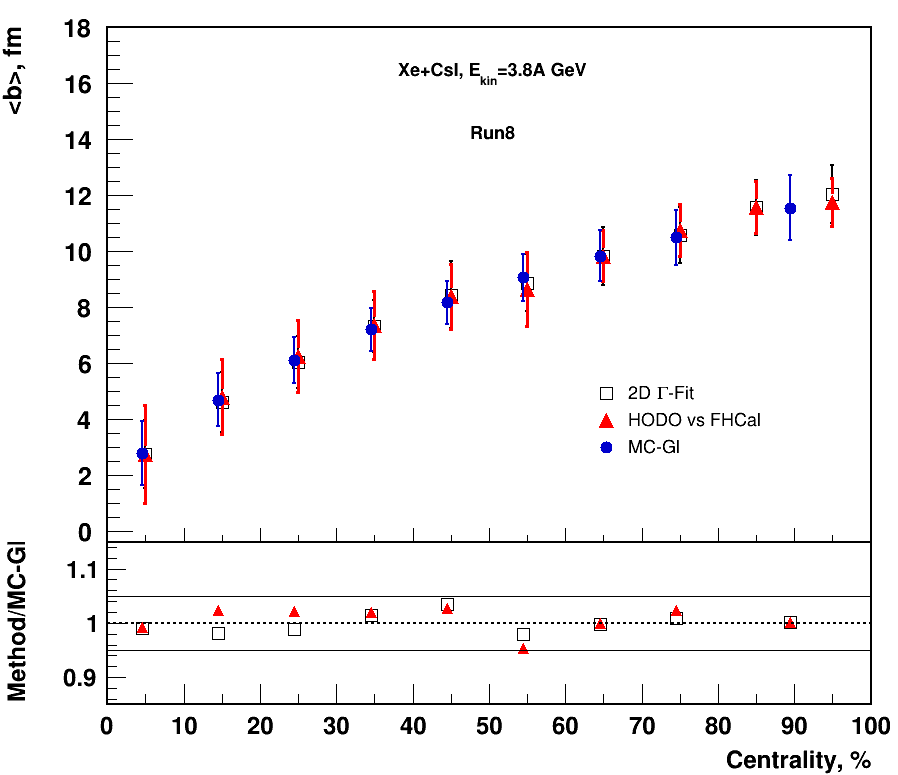}
\caption{Dependence of the mean impact parameter on centrality for various methods.}
\label{fig:impDistMean}
\end{figure}

Figure \ref{fig:impDist} presents a comparison of impact parameter distributions for three centrality classes(0-10\%, 30-40\%, and 60-70\%) obtained by different methods(MC-Glauber, two-dimensional methods with (\(E_{\text{spec}}\), \(N_{\text{hits}}\)) and only signals from the forward detectors). Figure \ref{fig:impDistMean} shows the dependence of the mean impact parameter on centrality. The obtained results indicate that all considered methods are in good agreement — differences do not exceed 5\% across the entire centrality range.

\clearpage

\section{Conclusion}

In this work, new approaches for determining centrality in nucleus-nucleus collisions based on the direct reconstruction method are proposed. The developed modification of the one-dimensional method allows accounting for calculated efficiency. Using charged-particle multiplicity to determine centrality, we obtained the event registration efficiency as a function of impact parameter with Bayesian approach. This allow more accurate comparison of model calculations with experimental data.

The two-dimensional method, based on the combined analysis of spectator energy from the forward hadron calorimeter FHCal and track hit count, allows reconstruction of the impact parameter using a two-dimensional gamma distribution. This approach may be useful in studies of net-proton multiplicity fluctuations, as it can suppress the autocorrelation effect.

The obtained results also formed the basis of a centrality determination method using the quartz hodoscope and forward hadron calorimeter FHCal. This approach can suppress the autocorrelation effect when analyzing proton multiplicity fluctuations in nucleus-nucleus collisions. The performed comparison of the dependence of the mean impact parameter on centrality obtained using the proposed approaches and the classical Monte Carlo Glauber method demonstrates good agreement across the entire centrality range, with differences not exceeding 5\%. The results confirm the reliability and mutual consistency of the developed methods which can be used to determine centrality in the BM@N experiment.

\acknowledgments{The authors express their gratitude to the BM@N collaboration for their contributions to data collection, and to the Accelerator Division of the Laboratory of High Energy Physics (LHEP), JINR.}

\conflictsofinterest{The authors declare no conflicts of interest.}

\funding{This research received no external funding.}

\institutionalreview{Not applicable.}

\informedconsent{Not applicable.}

\dataavailability{The data presented in this study are available on request from the corresponding authors.}

\printbibliography

\end{document}